# A New Method for Identifying Synthetic Speech Fakes based on Information-Geometric Superposed Vowel Evaluation: Part 1. Moraic Syllabary (Japanese)

Yusei TAMURA
Graduate School of Interdisciplinary Information Studies
The University of Tokyo
tamura-yusei@g.ecc.u-tokyo.ac.jp

Shigekazu ISHIHARA
Department of Psychology, Faculty of Health and Wellness Sciences
Hiroshima International University
i-shige@hirokoku-u.ac.jp

Ken ITO
Interfaculty Initiative in Information Studies
The University of Tokyo
itosec@iii.u-tokyo.ac.jp

## Abstract

This paper explains the principles and provides examples of a new method for distinguishing between FAKE human speech synthesized by generative AI and natural speech. Since synthetic speech is generated based on information from a limited set of training spectra, the variety of vowels—which are key to identifying individuals—is limited. In contrast, natural speech exhibits a more diverse distribution of vowel spectra due to the flexibility of the human articulatory organ. In this paper, using Japanese—a Syllabary limited to five vowel phonemes, each of which corresponds one-to-one with a specific sound—as an example, we outline a method for distinguishing between synthetic and natural speech reading the same text by analyzing the spectral distributions. If we normalize the spectra of speech sounds and regard them as probability density functions for the frequency bands received by the hair cells of human cochlea, and evaluate the distance between spectra using the Wasserstein metric, the Wasserstein distances between the vowels of synthetic speech are short. By preserving this distance and performing a topological mapping using persistent homology, the spectral probability density functions of synthetic and natural speech can be decomposed into clusters.

# 1 Introduction

Deepfakes have become a social issue and are used in various crimes, such as investment scams or kidnapping [1][2]. According to (Japanese) statistics, 90 percent of the public is aware of their existence and that it is difficult to distinguish them solely by listening [3]. However, only about 20 to 40 percent of the public is aware that synthetic speech can be clearly distinguished using appropriate methods [4], while the majority mistakenly believe otherwise. In Japan, these misconceptions are widespread, even among cabinet ministers and members of the Diet.

In this paper, we explain the principles and provide practical examples of a new method for clearly distinguishing AI-generated fake human voices from natural spoken language, building on our previous research into musical instrument timbres [5].

Synthetic speech learns from a limited set of training spectra to imitate the speech of a specific individual. Since it is based on information from a finite set of vowel spectra, its variety is intrinsically limited. In contrast, natural speech exhibits much greater variety because the flexibility of the human vocal tract and the redundancy of spoken-language vowels allow a single vowel to correspond to many different acoustic waveforms.

This paper provides an overview of methods for detecting deepfakes in Japanese; other languages, such as English, will be discussed in detail in subsequent papers. Japanese is regarded as moraic syllabary and its vowel letters are limited to five— "A, I, U, E, O"—and each corresponds to a specific moraic syllable one to one.

Using Japanese example sentences, by extracting the spectra of each vowel mora in both synthesized and natural speech as they read aloud the same text, and normalizing the spectra by dividing by their integral, it is possible to interpret these spectra as probability density functions for the frequency bands received by human hair cells in the cochlea (probabilistic spectroscopy).

When the distances between spectra are evaluated using the Wasserstein metric—employed in probabilistic optimal transport problems—the Wasserstein distances between synthetic vowels, that conform to a limited set of training spectra, are found to be short. By preserving this distance relationship and mapping the spectra onto a topological space using persistent homology, the spectral probability density functions of synthetic and natural speech can be decomposed into distinct clusters.

# 2 Theory

In the human auditory system, physical vibrations in the air are converted into traveling waves within the cochlea at the peripheral level; these waves are then decomposed into a frequency spectrum by a network of hair cells and transmitted to the central nervous system—specifically, the brain—via the cochlear nerve. It is important to note that this process is not a static Fourier transform, but rather a frequency decomposition of traveling waves.

It is not easy to replicate this with mechanical systems; instead, by performing a Fourier transform on the physical sound waves captured within a given time period and analyzing their spectrum, the system achieves highly intelligent acoustic information processing.

If we consider the cochlear nerve to be a detector that senses partial waves belonging to specific frequency bands, then the Fourier spectrum of a physical sound wave can be thought of as representing the probability that each frequency component will be perceived and detected by the human central nervous system within a given time interval.

For an arbitrary speech signal $v(t)$, consider its Fourier transform

$$\hat{v}(f) = \int_{-\infty}^{\infty} v(t)\, \mathrm{e}^{-\mathrm{i}2\pi f t}\, \mathrm{d}t\,. \tag{1}$$

Next, let

$$V = \int_{-\infty}^{\infty} \hat{v}(f)\, \mathrm{d}f \tag{2}$$

be the integral of $\hat{v}(f)$ over the entire frequency domain, and define the normalized spectrum

$$\mathcal{F}(f) \equiv \frac{\hat{v}(f)}{V}. \tag{3}$$

We interpret $\mathcal{F}(f)$ as a probability density function representing the probability that an observer perceives each frequency component during a short time interval (Stochastic spectroscopy).

Let's consider the multiple spectra obtained in this way and the "distance" between them. This should serve as a metric for quantifying the "difference" or "similarity" between two timbres. Various measures of spectral "similarity" can be considered, such as distance in the Euclidean sense, Manhattan distance (which is the sum of the absolute differences in frequency), or KL divergence using the Fisher metric [5]. Among these, the Wasserstein distance—which is used to solve the optimal transport problem—is considered suitable for evaluating the perceptual "similarity of timbre" because it treats spectra that differ only slightly in pitch, as if produced by the same speaker, as being close to one another.

Now, for the spectral probability density functions $\mathcal{F}_1(f)$ and $\mathcal{F}_2(f)$, consider the cumulative spectral density functions on the positive frequency axis $[0, \infty)$:

$$\mathcal{M}_1(f) = \int_0^f \mathcal{F}_1(f')\, \mathrm{d}f'\,, \tag{4a}$$

$$\mathcal{M}_2(f) = \int_0^f \mathcal{F}_2(f')\, \mathrm{d}f'\,. \tag{4b}$$

The one-dimensional Wasserstein distance $W_{1,2}$ [6][7] can then be defined as

$$W_{1,2} = W_1(\mathcal{F}_1, \mathcal{F}_2) = \int_0^{\infty} |\mathcal{M}_1(f) - \mathcal{M}_2(f)|\, \mathrm{d}f\,. \tag{5a}$$

For numerical computation, it is often more convenient to use the inverse cumulative spectral density functions, giving the equivalent expression

$$W_{1,2} = \int_0^1 \left|\mathcal{M}_1{}^{-1}(\alpha) - \mathcal{M}_2{}^{-1}(\alpha)\right| \mathrm{d}\alpha\,. \tag{5b}$$

As is evident from Eq. (5a), the Wasserstein distance defined in this manner assigns only a small value to slight changes in the positions of peaks along the frequency axis. Consequently, translational shifts of the spectrum along the frequency axis—that is, changes in pitch—are regarded as only minor differences. For this reason, we consider the Wasserstein distance to be well suited for quantifying the perceptual distance between speech spectra.

Suppose that the spectral probability density functions obtained from a set of speech samples belonging to the same class are arranged as a set of spectral vectors, denoted by $\Psi$. Let $W_{ij}$ denote the Wasserstein distance between the $i$-th and $j$-th spectral vectors $\psi_i, \psi_j \in \Psi$. Arranging these pairwise distances yields the Wasserstein distance matrix

$$D = \begin{pmatrix} W_{00} & W_{01} & W_{02} & \cdots & W_{0n} \\ W_{10} & W_{11} & W_{12} & \cdots & W_{1n} \\ W_{20} & W_{21} & W_{22} & \cdots & W_{2n} \\ \vdots & \vdots & \vdots & \ddots & \vdots \\ W_{n0} & W_{n1} & W_{n2} & \cdots & W_{nn} \end{pmatrix}. \tag{6a}$$

By definition, $W_{ii} = 0$, and therefore

$$D = \begin{pmatrix} 0 & W_{01} & W_{02} & \cdots & W_{0n} \\ W_{10} & 0 & W_{12} & \cdots & W_{1n} \\ W_{20} & W_{21} & 0 & \cdots & W_{2n} \\ \vdots & \vdots & \vdots & \ddots & \vdots \\ W_{n0} & W_{n1} & W_{n2} & \cdots & 0 \end{pmatrix}, \tag{6b}$$

which is an $n \times n$ real symmetric matrix. Hence it is diagonalizable, yielding the eigenvalues

$$\lambda_k \qquad (1 \leq k \leq n) \tag{7}$$

together with the corresponding eigenvectors $\phi_k$.

The number of vectors in $\Psi$ is determined by the number of samples. However, because a sample set typically consists of only $10^1$–$10^2$ samples, it is generally difficult to interpret such a high-dimensional space directly. In the following, we therefore consider the orthogonal projection onto a subspace of dimension at most three spanned by the eigenvectors corresponding to the largest eigenvalues, and discuss the results together with their visualization.

## 3 Method

Below is a brief overview of a specific routine for identifying deepfake audio in Japanese audio files.

Deepfakes generally use the voices of people who are well-known to the public. Cases in which the likenesses or voices of celebrities are used or repurposed without permission—for criminal activities such as fraud, or for abusive purposes during public elections—are of particular concern.

Therefore, in the following,

1. Audio documents suspected of being deepfakes
   ($K$: **Knock-off**)

and

2. Documents confirmed to be natural speech from the person whose voice was used without permission as training data
   ($O$: **Original**)

will be used to conduct the verification.

First, as a preliminary experiment, we obtained permission to use a voice sample from Nobuo Gohara, a prominent lawyer in Japan's online community and a friend of the authors, and compared natural speech with synthesized speech for the five Japanese vowel morae.

For Gohara's natural speech samples $O_G$ and the synthesized speech samples $K_G$ of the same sentences generated by a system trained on those samples, the five Japanese vowel morae, **A, I, U, E,** and **O**, were extracted. Their normalized spectra were then represented as collections of spectral probability density functions, denoted by $\Psi_{O_G}$ and $\Psi_{K_G}$, respectively. From these, the corresponding $5 \times 5$ Wasserstein distance matrices, $D_{O_G}$ and $D_{K_G}$, were constructed.

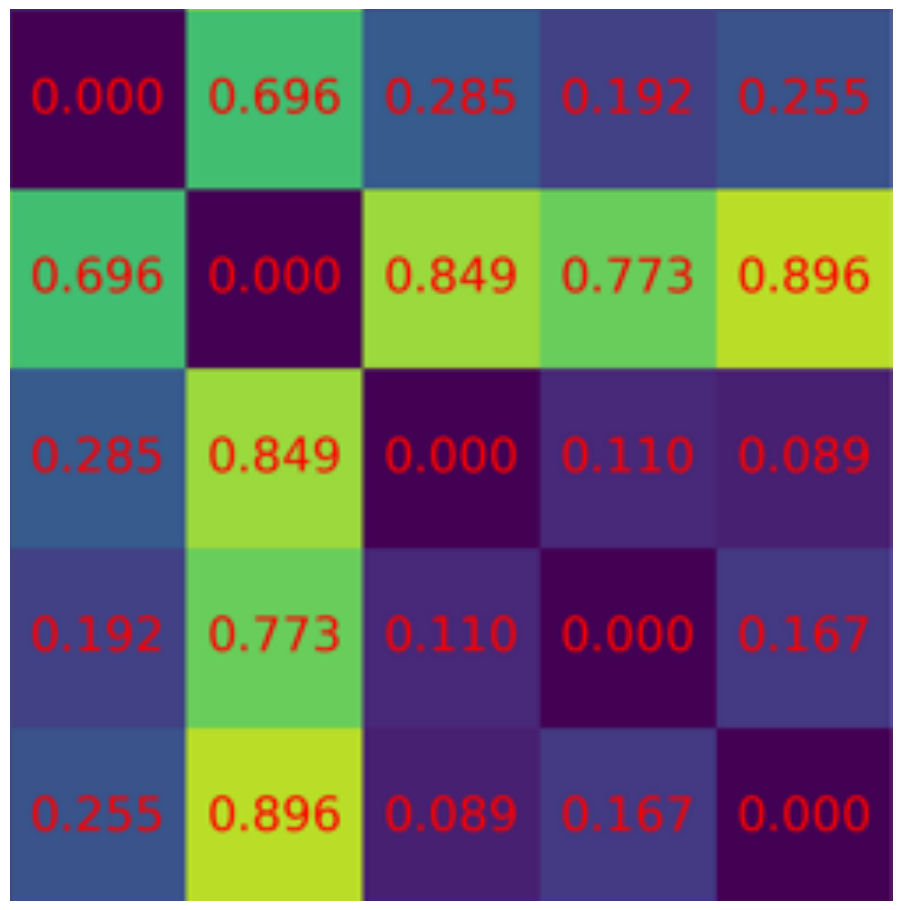

Fig. 1-1 Wasserstein Distance Matrix $D_{O_G}$ for the Five Natural Vowel Morae

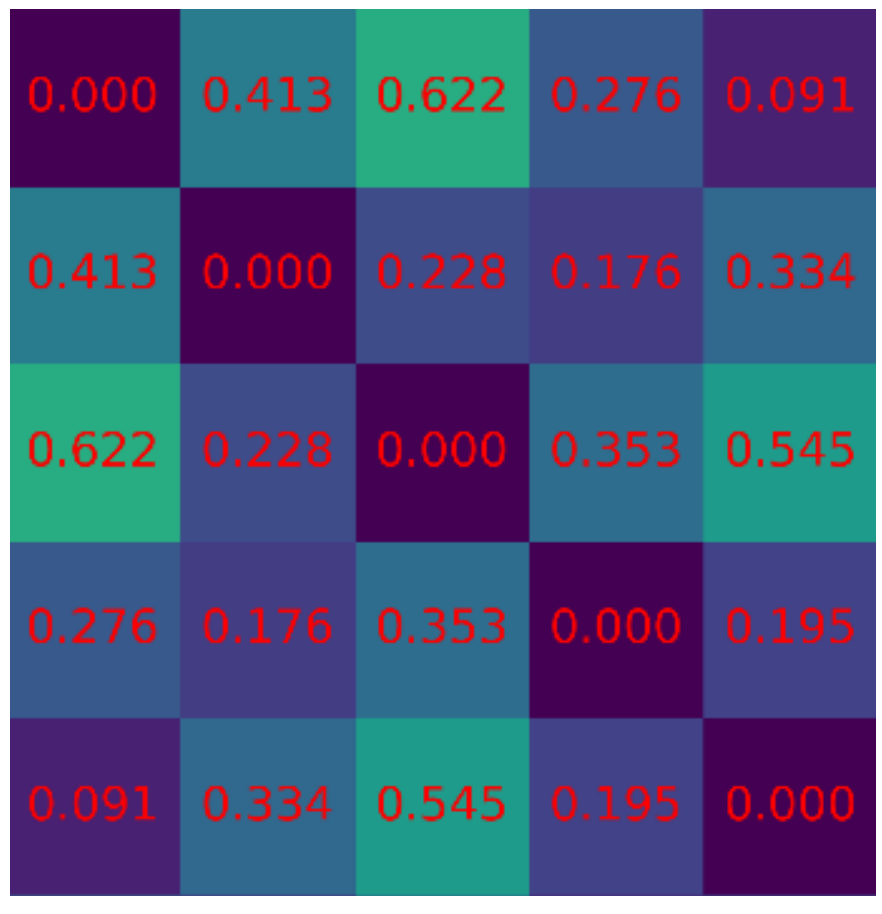

Fig. 1-2 Wasserstein Distance Matrix $D_{K_G}$ for the Five Vowel Morae of Synthetic Speech

In practice, to treat the two datasets on an equal footing, the calculations are performed using the augmented distance matrix $\Psi_{OK_G}$, obtained by concatenating $\Psi_{O_G}$ and $\Psi_{K_G}$. The results presented here are normalized by the maximum value.

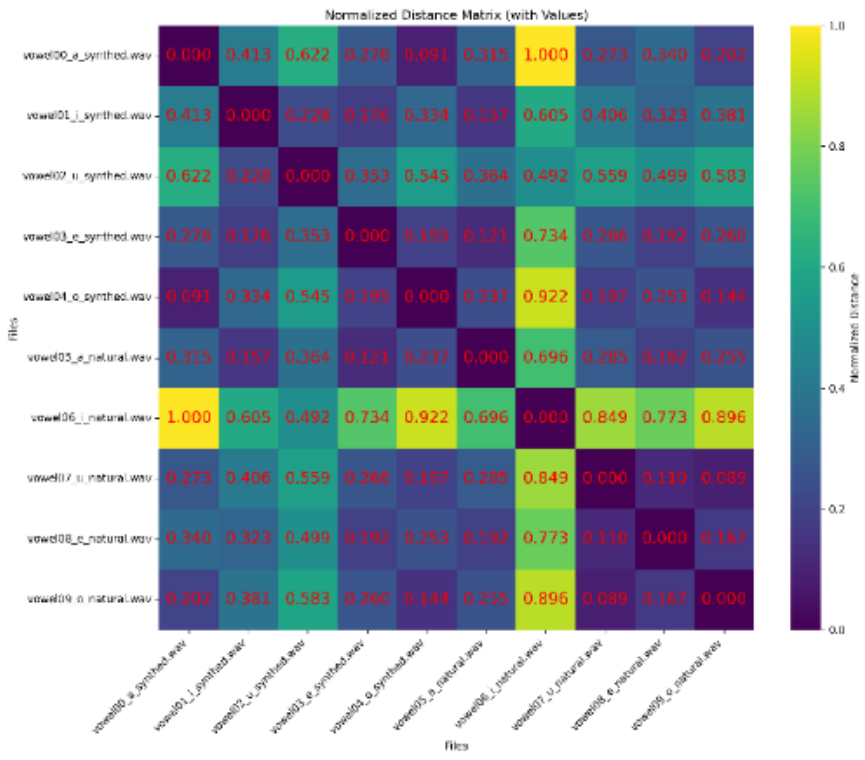

Fig. 1-3 Expanded Wasserstein Distance Matrix $D_{OK_G}$ for the Five Vowel Morae

To make the numerical values visually clear, the data has been colored using a scale where 0 is dark blue and 1 is yellow. Even a quick glance at the three matrices in Fig. 1 reveals that the Wasserstein distance between natural vowel morae spectra is larger than that of synthesized speech. The mean value of the (off-diagonal components of the) normalized Wasserstein distance matrix is ** for natural speech and ** for synthesized speech.

Fig. 2 shows the timbre spectra mapped onto phase space using the persistent homology method [8], while preserving these Wasserstein distances.

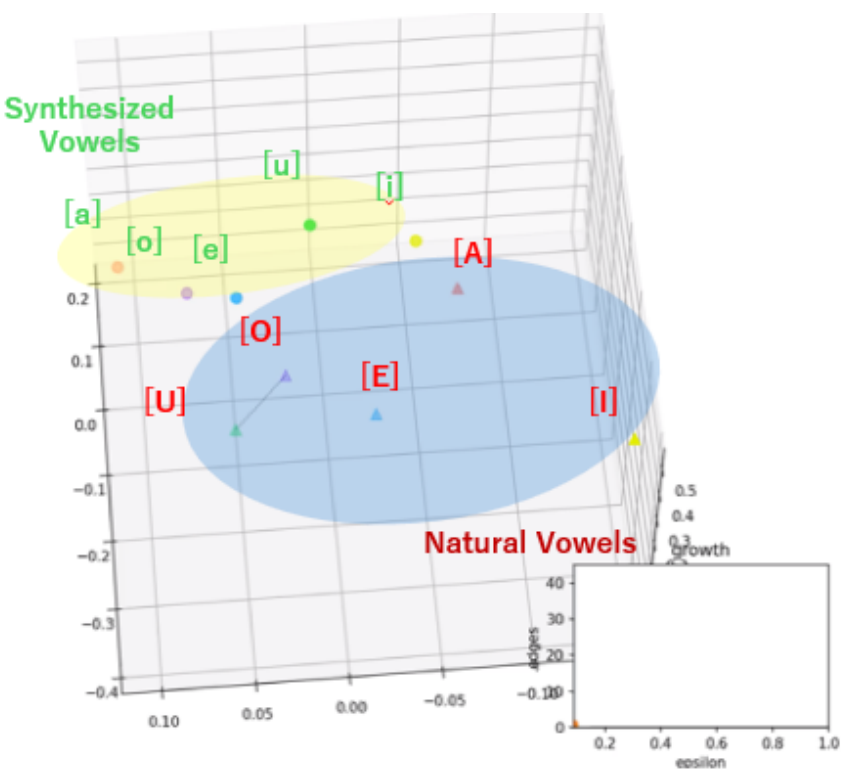

Fig. 2 Topological Mapping of Natural and Synthetic Vowel Sounds Using Persistent Homology

When the spectrum is mapped onto a topological space while preserving the Wasserstein distance, it is observed that synthetic vowel morae are localized in a relatively narrow region, whereas vowel morae in natural speech are distributed over a wide range, allowing the two to be clearly distinguished.

Next, we will examine how to handle documents suspected of being fake within the overall dataset. This is done with an eye toward extending and expanding this method to languages other than Japanese—such as English—where syllables and letters do not have a one-to-one correspondence, as they do in Japanese.

Let $\chi(f)$ denote the normalized audio spectrum obtained by applying the Fourier transform to the entire audio document $X(t)$, which is suspected of being FAKE. We propose examining the frequency distribution of the vowels and consonants contained in this data, generating synthetic speech that is identical or similar to it using a system trained on the voice of the same speaker as the training data for the suspicious document, and then comparing the two sets. As model cases, we generated five test sentences using a large-scale language model, each containing roughly the same number of each of the five Japanese vowels. Specifically, we created the following five Japanese sentences.

Table 1. Japanese test texts with quasi-equal distribution of 5 vowel letters

| Japanese Sample texts | Translation |
|---|---|
| **1 Aoi uwagi no oi no negai o kanaeyo** | **Grant blue-jacketed nephew's wish.** |
| **2 Aoi eki e to utau ojisan aruiteku** | **Singing uncle walks to blue station.** |
| **3 Ai o utaeba oi mo egao de kotaeso** | **Sing love, nephew smiles back.** |
| **4 Akai ienami ou ehagaki o atumeyo** | **Collect postcards chasing red houses.** |
| **5 Aoi edamame ou yo ni tanomi ajiwao** | **Order and enjoy green peas.** |

As is evident immediately from Table 1, the Japanese test texts do not convey a coherent meaning. However, the rate of vowel occurrences is controlled, making it possible to utter them naturally as Japanese sentences.

The speech synthesis system is trained separately using training audio from the same speaker, with each recording containing all of the five Japanese vowels.

Both this system and the original speaker were made to utter the five test sentences shown in Table 1, and the entire recording was subjected to a Fourier transform to obtain normalized spectra.

Since these five sentences contain five vowels in controlled proportions (weights), the Fourier transforms of each audio document yield a composite spectrum in which the contributions of the various vowels are roughly equal.

The expanded distance matrix obtained in this way is shown in Fig. 3.

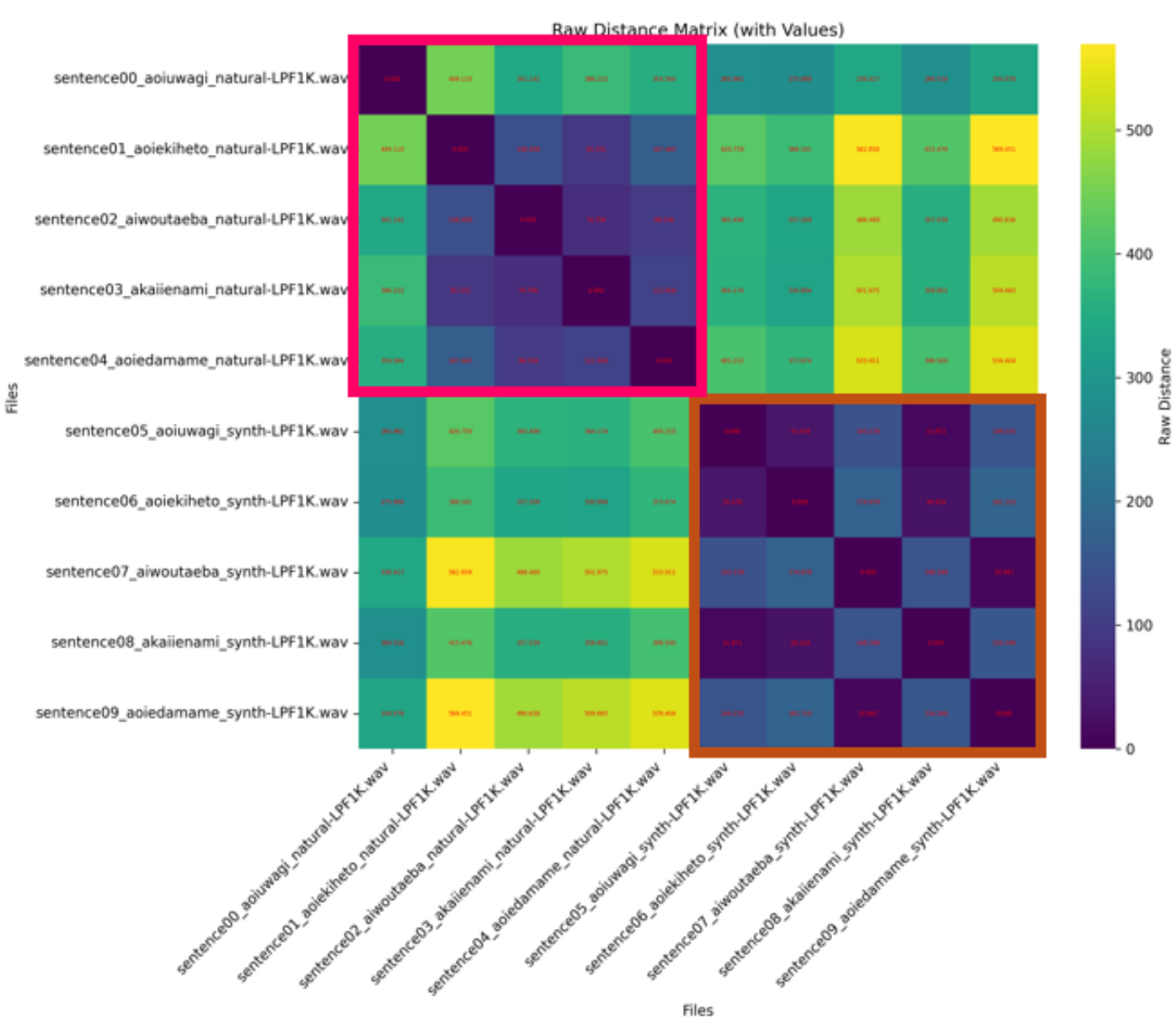

Fig. 3 Expanded Wasserstein distance matrices for five test sentences with controlled rate of vowel occurrence, using natural / synthesized speech.

## 4 Results

As shown in Fig. 3, for a sample document containing five sentences that are completely identical when viewed as text, the Wasserstein distance between synthesized samples is significantly shorter than that between natural samples.

The overall distribution can be mapped using the persistent homology method, which places documents as points in a topological space while preserving the Wasserstein distance between spectra. Let's look at the results for the example above.

For audio documents consisting of entire sentences, it was confirmed that the spectra of synthesized speech are localized in phase space, whereas those of natural speech are dissociated. This trend was observed in all similar experiments.

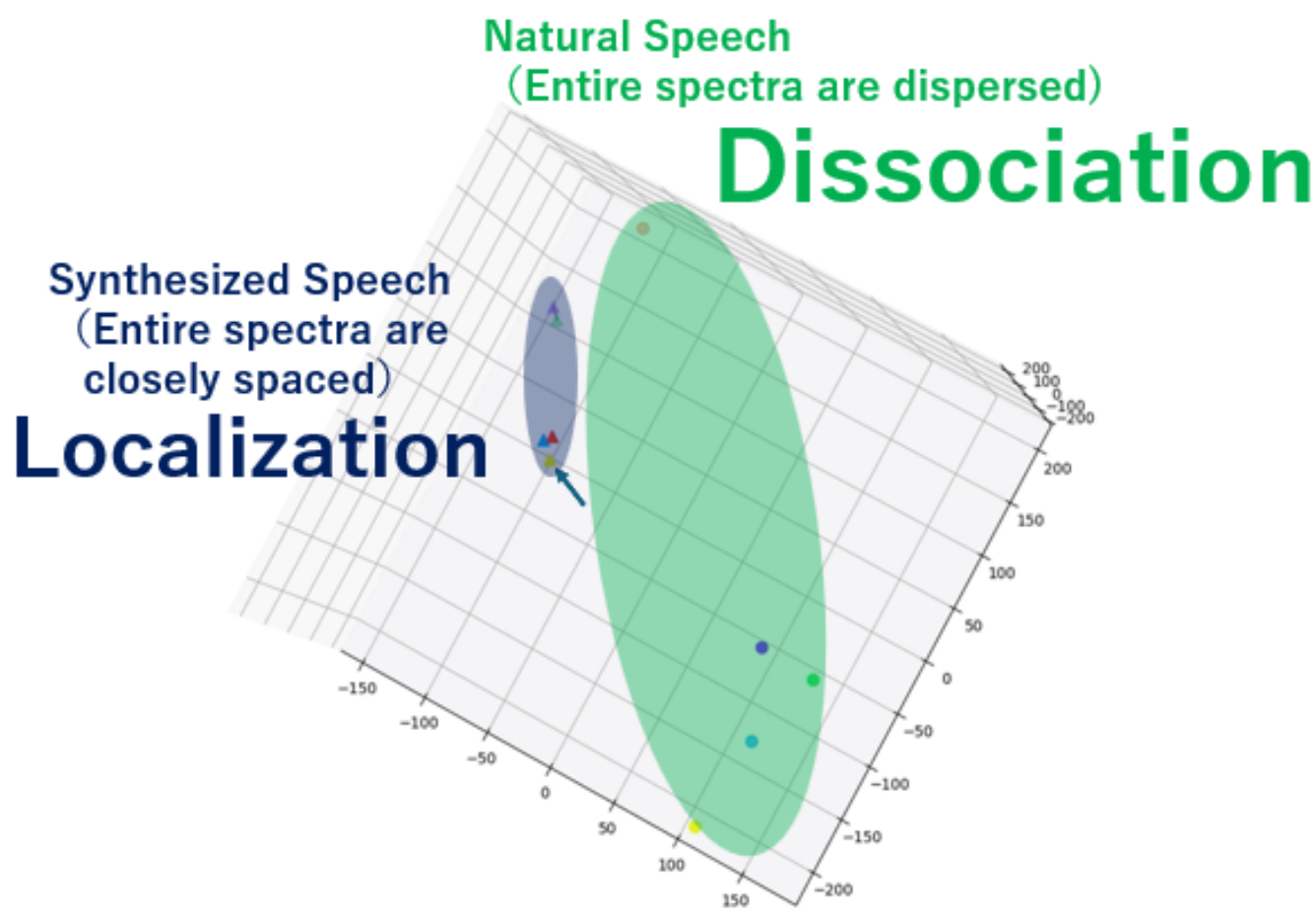

Fig. 4 A topological mapping of the normalized spectra of entire sentences — both natural and synthetic speech — using persistent homology whilst preserving the Wasserstein distance.

## 5 Discussion: Toward Expanding into Arbitrary Syllabary, English and Other Languages

Previous examples of the application of topological mapping using persistent homology to speech-related tasks include Bonafos et al. (2023) for vowel classification and Liu et al. (2017) for musical audio signal processing. The former solves three prediction problems—speaker gender, vowel type, and individual identity—based on vowel audio recordings, while the latter proposes incorporating a topological summary of audio data into a convolutional neural network (CNN) for signal processing. Although both are significant contributions, they differ from the objective of distinguishing features in audio data generated by generative AI systems from those of natural speech.

In the case of Japanese, a moraic syllabary, since there are only five vowel letters and there is a formal one-to-one correspondence between syllables and pronunciation, the approach taken in this paper proved effective.

In contrast, the number of vowels in English and any other arbitrary syllabary is in itself a matter of debate; while it is officially considered to have about 20 vowels, the correspondence between vowels and the written representation of syllables is not straightforward (arbitrary).

Therefore, when using an LLM to analyze audio documents suspected of being fake in order to determine whether they sound natural or not, it may be possible to conduct the analysis after converting them to a phonemic notation system, such as ARPAbet[12], which was developed in the United States during the Cold War.

Suppose we have a sample text "Contextual classification using character n-grams in phonemic notation", it would be written in ARPAbet as

[K AA N T EH K S CH UW AH L K L AE S AH F AH K EY SH AH N Y UW Z IH NG K EH R IH K T ER EH N G R AE M Z IH N F OW N IY M IH K N OW T EY SH AH N].

Using phonemic notation sequences such as ARPAbet, it is possible to construct alternative sentence examples with the same vowel occurrence rate and examine the results of phonetic analysis using those examples. For the example mentioned above, an alternative sentence such as "An elegant tutor showed that using raw data can teach us each music notation" can be prepared (refer to Appendix).

Examples of analyses for arbitrary syllabary text using these methods are described in detail in the Paper Part 2, which will be published soon.

# 6 Concluding remarks

By treating the acoustic spectra of vowels and sentence documents as probability distribution functions and evaluating the similarity of phonemes using their Wasserstein distances, we show that Topological Mapping enables a clear distinction between fake and natural speech.

Since this approach relies on LLMs, the congruence between written symbols and speech sound is crucial; in this paper, we apply this method to moraic syllabary, namely Japanese, where that relationship is relatively simple. We plan to discuss its application to arbitrary syllabary in the subsequent paper.

# Acknowledgement

The authors would gratefully show best thanks to Dr. Prof. Shun-ichi AMARI for his warm advice. We also send cordial thanks to DynaxT Co. and our seminar students at College of Arts and Sciences, The University of Tokyo.

## Appendix

When “Contextual classification using character n-grams in phonemic notation” is written using ARPAbet and the frequency of vowel occurrences is organized, the result is as follows:

AH : 5 (contextu[AH]l, classific[AH]ti[AH]n, not[AH]ti[AH]n)

EH : 3 (cont[EH]xtual, char[EH]cter, [EH]n-grams)

IH : 3 (us[IH]ng, char[IH]cter, [IH]n)

AA : 1 (c[AA]ntextual)

AE : 2 (cl[AE]ssification, n-gr[AE]ms)

EY : 2 (classific[EY]tion, not[EY]tion)

UW : 2 (context[UW]al, [UW]sing)

ER : 1 (charact[ER])

OW : 2 (ph[OW]nemic, n[OW]tation)

IY : 1 (phon[IY]mic)

In contrast, the vowel frequency for “An elegant tutor showed that using raw data can teach us each music notation.” is:

AH (5): elegant (2), data (1), can (1), us (1) = 5

EH (3): elegant (1), text (1), [cancelled out by contextual adjustment before “notation”] = 3

IH (3): using (1), music (1), adjustment before “notation” = 3

AA (1): raw (1) = 1

AE (2): An (1), that (1) = 2

EY (2): data(1), notation(1) = 2

UW (2): tutor(1), music(1) = 2

ER (1): tutor(1) = 1

OW (2): showed(1), notation(1) = 2

IY (1): each(1) = 1

The results match.